\documentclass{article}
\usepackage{graphicx} 
\usepackage{xcolor} 
\usepackage[utf8]{inputenc} 
\usepackage[version=4]{mhchem} 
\usepackage{chemfig} 
\usepackage[export]{adjustbox}
\usepackage{amssymb}
\usepackage{comment}
\graphicspath{{C:\Users\rshojaei\Desktop}}

\newcommand{\abstractnamefont}{}
\newcommand{\abstracttextfont}{}
\newcommand{\eq}[1]{\begin{equation}#1\end{equation}}

\title{Counting Chemical Isomers with Multivariate Generating Functions}
\author{Rana Shojaei and Thilo Gross} 
\date{%
    \normalsize
    Helmholtz Institute for Functional Marine Biodiversity (HIFMB), Oldenburg, Germany\\
    Alfred Wegener Institute, Helmholtz Centre for Marine and Polar Research, Bremerhaven, Germany\\
    Carl von Ossietzky University, Institute for Chemistry and Biology of the Marine Environment (ICBM)-Oldenburg, Germany\\[2ex]%
    \today}
\begin{document}
\maketitle

\begin{center}
    {\abstractnamefont Abstract}  
\end{center}
\vspace{0.5em} 

\noindent
\begin{center}
    \begin{minipage}{0.8\textwidth} 
        \abstracttextfont Counting the number of isomers of a chemical molecule is one of the formative problems of graph theory. However, recent progress has been slow, and the problem has largely been ignored in modern network science. Here we provide an introduction to the mathematics of counting network structures and then use it to derive results for two new classes of molecules. In contrast to previously studied examples, these classes take additional chemical complexity into account and thus require the use of multi-variate generating functions. The results illustrate the elegance of counting theory, highlighting it as an important tool that should receive more attention in network science.  
    \end{minipage}
\end{center} 
\vspace{1em} 


\vspace{2em} 


\section{Introduction} 
%
%
Modern network science is grounded in graph theory, the mathematical study of networks, which famously dates back to Euler's solution of the K\"{o}nigsberg bridge problem. However, the usage of term `graph' to describe a network is younger and can be traced back to Sylvester's work on the number of chemical isomers with a given sum formula \cite{sylvester1878application}. 

In chemistry, the sum formula specifies the number of atoms of different chemical species that are contained in a given molecule. For instance, the sum formula of water, H$_2$O, shows that a molecule of water contains one oxygen and two hydrogen atoms. Within a molecule, atoms are held together by covalent bonds, and in stable, electrically neutral molecules each atom participates in a characteristic number of such bonds, depending on the species of atom. Hence, when a molecule is interpreted as a network, where the atoms are nodes and chemical bonds are links, carbon atoms have to correspond to nodes of degree four, oxygen atoms correspond to nodes of degree two, and hydrogen atoms correspond to nodes of degree one. 

To ensure reasonable stability of molecules some additional constraints are necessary. In the context of the present paper the most important of these constraints is that any molecule containing long chains of oxygen atoms is unstable and would rapidly decay. 

\begin{figure}[h]
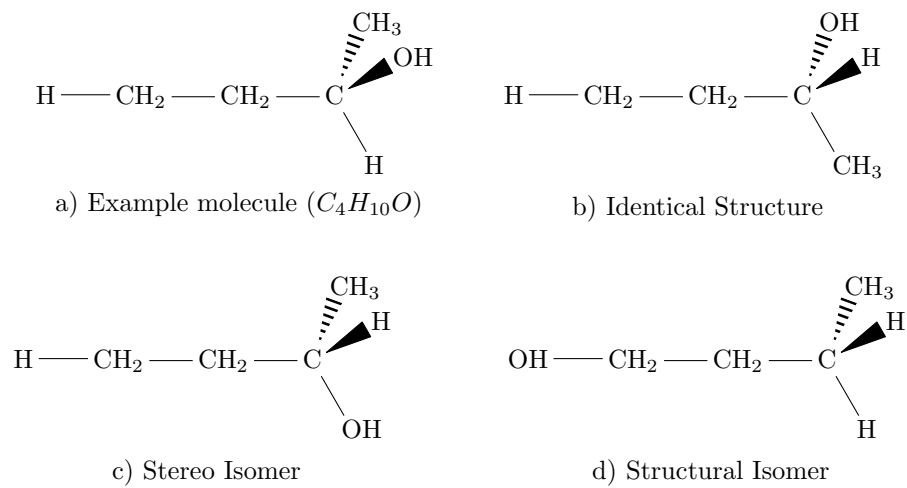
 
\centering
   \chemname{%
    \chemfig{%
      CH_{2}(-[:180]H)-[:0]CH_{2}-[:0]C(<[:30]OH)(-[:300]H)(<:[:70]CH_{3})
    }
  }{a) Example molecule ($C_{4}H_{10}O$)}
  \qquad
   \chemname{%
    \chemfig{%
     CH_{2}(-[:180]H)-[:0]CH_{2}-[:0]C(<[:30]H)(-[:300]CH_{3})(<:[:70]OH)
    }
  }{b) Identical Structure}

  \vspace{.7cm}

  \chemname{%
    \chemfig{%
      CH_{2}(-[:180]H)-[:0]CH_{2}-[:0]C(<[:30]H)(-[:300]OH)(<:[:70]CH_{3})
    }
  }{c) Stereo Isomer}
  \qquad
  \hspace{.5cm}
  \chemname{%
    \chemfig{%
      CH_{2}(-[:180]OH)-[:0]CH_{2}-[:0]C(<[:30]H)(-[:300]H)(<:[:70]CH_{3})
    }
  }{d) Structural Isomer}
  \caption{Illustration of different types of isomers. Chemical diagrams illustrate the structure of molecules consisting of hydrogen (H), oxygen (O), and carbon atoms (C). The bonds between these atoms are in the plane of the figure (lines) point toward the reader (filled triangle) or away from the reader (dashed triangle). All carbon atoms have degree 4, but following chemical convention, the links in $-CH_{2}$ and $-CH_{3}$  are omitted to reduce clutter. Likewise, the link between oxygen and hydrogen is not shown as a bond. The figure illustrates that (a) and (b) are identical structures as they differ only by rotation. Molecule (c) is a stereo-isomer of (a) because it is topologically identical but related to (a) by a mirror-symmetry, i.e. (a) cannot be transformed into (c) by rotations alone. Finally (d) is a structural isomer of (a) as it has the same sum formula but a different topology.}
    \label{fig: isomer examples}
\end{figure} 

For the water molecule, the rules permit only a single configuration, $H_{2}O$, but in larger organic molecules, a single sum formula can correspond to multiple and possibly very many distinct network configurations. In chemistry, the different configurations corresponding to a single sum formula are called isomers. To count how many isomers exist that are consistent with a given sum formula, we must first define which configurations we consider to be the same and which ones we consider to be different. One option is to consider two molecules to be identical isomers if they have the same topology, another possible choice is to consider two molecules as identical if their atoms can be oriented in the same way in three-dimensional space. In the former case, the bonds of an atom are completely interchangeable, and we are counting so-called \emph{structural isomers}. In the latter case, the bonds can only be interchanged by geometrical rotations, and we are counting \emph{stereo-isomers} (Fig.~(\ref{fig: isomer examples})).

In principle, one can determine the number of configurations by explicit enumeration, i.e.~by iterating through all ways to realize the desired degree sequence. However, this approach becomes infeasible both due to the number of possible configurations and the difficulty of deciding whether two configurations are isomers or identical structures~\cite{read_graph_1977}. Moreover, note that counting the isomers of a given degree sequence and arriving at a result that is greater than zero demonstrates that the degree sequence can be realized as a graph. Hence the problem of deciding whether a given degree sequence is graphical, which is a difficult problem in its own right~\cite{hakimi_realizability_1962}, is solved as a byproduct of the isomer counting problem. This suggests that there is no easy universal solution to the isomer counting problem. However,  progress can be made for broad classes of structures.    
   
In mathematics, the counting of configurations of systems falls into the realm of enumeration theory, which dates back to the works of Cayley~\cite{cayley1874lvii, 1875ancayleyalytical}, Redfield~\cite{redfield1927theory}, and, P\'{o}lya~\cite{polya1937kombinatorische}. A convenient method for counting configurations in the presence of symmetries was proposed by Redfield~\cite{redfield1927theory} and later independently rediscovered in a more convenient form by P\'{o}lya~\cite{polya1937kombinatorische} and extended by Read~\cite{read}. P\'{o}lya's theorem uses generating functions to represent the symmetries of a network, which allows, for example, to very efficiently calculate the number of ways in which the nodes on an unlabeled network can be colorized with a given number of colors. 
Later Harary extended P\'{o}lya's theory to count the number of branching network structures that obey certain constraints~\cite{harary2018graph}.

In the past, Harary's extension has been used to derive counting functions for different classes of molecules~\cite{wang2003enumeration, faulon2005enumerating}. However, these examples focus on molecules consisting of only carbon and hydrogen and possibly one other atom. In such molecules, the mathematics only needs to account for the molecule's carbon backbone, for which univariate generating functions are sufficient. Beyond these works in the early 2000s, the recent attention has mostly focused on the explicit enumeration of molecular structures using large-scale numerical efforts (e.g.~\cite{ruddigkeit_enumeration_2012}). 

Presently the interest in the number of isomers of molecules is being rekindled due to the rise of microbial ecology. For unicellular organisms the diversity of chemical isomers may be an important factor limiting metabolic rates. This is particularly interesting in the oceans where an enormous diversity of isomers is found~\cite{hawkes2018extreme, catala2021marine}. This diversity may prove to be the major factor that limits the rate at which bacteria can utilize dissolved organic carbon. It might thus be this diversity of molecules which limits the re-emission of carbon into the atmosphere in the form of CO$_2$. However, more work is needed to understand the sources and implications of molecular diversity in the oceans and other systems. 

Here we provide a gentle introduction to the counting of network structures, aimed at network scientists (Sec.~\ref{sec:Counting-Theory}). We then apply this approach to the counting of treelike molecules of carbon, oxygen ,and hydrogen, growing from a given, possibly cyclic, root. We extend previous mathematical work by considering more complex cases (including for example double and triple bonds), which necessitates the use of multivariate generating functions (Sec.~\ref{sec:GCF}). Finally we apply the approach to count the number of treelike molecules with naphthalene and benzene roots, which haven't been studied in this way (Sec.~\ref{sec:RingStructures}). We conclude by discussing the application to other classes of molecules and other potential applications in the context of network science.  


\section{Counting with Generating Functions \label{sec:Counting-Theory}}   
%
%
Let us start by revisiting P\'{o}lya's theory of counting and its subsequent extensions by Harary and others~\cite{harary2018graph}. Our aim here is to provide an intuitive approach to the theory to complement the mathematically rigorous formulations found in the literature.  

\subsection{Generating Function Basics} 
Most network scientists will be familiar with generating functions such as the degree generating function 
\eq{\label{eqGF}
G(x)= \sum p_k x^k,
}
where the $p_k$ are the elements of the degree distribution and $x$ is an abstract variable that is introduced for the purpose of encoding these elements in a continuous function.  

One way to think of generating functions is to consider them Taylor expansions in reverse. While the Taylor expansion transforms a continuous function into a set of coefficients, the formulation of a generating function turns a set of coefficients into a continuous function. Indeed, the Taylor expansion of a generating function recovers the sequence of coefficients, e.g.~$p_k$. Thus, the Taylor expansion and generating functions allow us to bridge between the worlds of continuous and discrete mathematics and thus bring the tools of the respective other world to bear on our problem. 

There is also a different way to interpret generating functions, which starts with the observation that the algebraic ring formed by multiplication and addition on the set of real numbers is congruent to the ring of Boolean operations on a set of discrete objects. What this means is we can read generating functions as descriptions of sets as follows:
\begin{itemize}
\item Read addition (+) as `OR'
\item Read multiplication ($\cdot$) as `AND' 
\item Interpret variables ($x$) as the representation of objects themselves (i.e.~not as variables that contain a number) 
\end{itemize}
Thus, for the degree distribution, the variable $x$ is no longer abstract but becomes a placeholder for the endpoint of a link (a stub) situated on the node. So $x$ now represents one link starting at a node, and $x^2=x\cdot x$ represents two links starting at a node. 

For example, a network may have the degree generating function
\eq{
G= 0.8 x + 0.2 x^4 ,
}
which we can read as 
\begin{center}\it
``If you pick a random node, you get a node with one link ($x$), with 80\% probability, OR you get a node with four links ($x^4$), with 20\% probability'',  
\end{center}
which is consistent with our usual reading of Eq.~(\ref{eqGF}).

In the example, we have interpreted numbers that appear in front of terms as probabilities, but using a different normalization, we can also interpret them as numbers. For example, if the network from the example has 10 nodes, we might write $8x+2x^4$ to say that there are 8 nodes of degree one and 2 nodes of degree four. When used in this way, we refer to generating functions as \emph{counting functions}.

\subsection{Simple Multivariate Counting} 
Formulating counting functions for sets of different types of objects leads to multivariate functions. For instance, a bag of marbles that contains 2 white marbles (w) and 3 grey marbles (g) could be described by the expression $w^2g^3$. In this interpretation, the function 
\eq{\label{eqGFexample}G(w,g)=1+w^2+3gw,} 
would mean that we have five bags in total: three that contain a white marble and a grey marble, one bag that contains two white marbles, and one bag that contains nothing ($w^0g^0=1$).

Once we have the generating function for our collection of marbles we can compute some numbers of interest directly from the generating function by using a combination of substitutions and derivatives. 

This is only of minor importance for our calculation below and many readers will already be familiar with the computation of a network's mean degree $z$ from the degree generating function as $z=G'(z)$. Let us nevertheless show some multivariate examples of simple counting problems without detailed explanations.

Form Eq.~(\ref{eqGFexample}) we can compute the total number of bags as 
\eq{
\label{eqCompFirst}
G(1,1)=5. 
}
Using roman indices to denote derivatives, we can write the total number of grey marbles as 
\eq{
G_{\rm g}(1,1) =3,
}
The number of bags that contain an odd number of white marbles is  
\eq{
\frac{G(1,1)-G(-1,1)}{2} = 3, 
}
And, the number of white marbles in bags that don't contain grey marbles is 
\eq{
\label{eqCompLast}
G_{\rm w}{(1,0)} = 2
}
Although we evaluated the quantities for the specific example of the counting function from Eq.~(\ref{eqGFexample}), the equations on the left-hand side compute the respective properties from arbitrary counting functions $G$. 

 \subsection{Network Coloring}
The example computations from Eqs.~(\ref{eqCompFirst}-\ref{eqCompLast}) are not very useful if we have direct access to a counting function as a polynomial, such as in Eq.~(\ref{eqGFexample}). However, computing properties of interest directly from the counting function becomes useful when we obtain counting functions as a result of calculations. In this case, we find the counting functions often in a more compact form that is tedious to expand into a polynomial, but can still be evaluated similarly to the examples above. 

To illustrate the derivation of a counting function, let's consider a very simple network consisting of a linked pair of nodes (o-o). Suppose we are interested in the number of configurations that we can create if we color the nodes white or grey. To start exploring this question we might start by stating that there are two nodes in the network:
\eq{
G=n^2,
}
where $n$ now represents a node. Next, we observe that each node can either be white OR grey which in terms of generating functions can be written as 
\eq{
\label{eqColorSubs}
n=w+g
}
and hence 
\eq{
\label{eqCompactEx}
G=(w+g)^2 
}
we could now expand this generating function to obtain the polynomial $w^2+wg+g^2$, but the point is, that we don't have to do this as many properties of interest can be computed directly from the compact form of Eq.~(\ref{eqCompactEx}) using approaches such as Eqs.~(\ref{eqCompFirst}-\ref{eqCompLast}).

We may count the total number of configurations that we can create by evaluating $G(1,1)=4$. Generalizing from the example, we can see that we would describe a network with $N$ nodes by $G=n^N$. If we color the network using $C$ colors, then the equivalent of Eq.~(\ref{eqColorSubs}) would equate $n$ with a sum over $C$ color variables. If we then set all these variables to one to count the configurations, we find $n=C$, and substituting this into the generating function gives the expected number of configurations $G=C^N$. This is hardly a surprising result, but seeing how this simple and intuitive result emerges from the counting function lays the foundation for the more complex counting theory that we now revisit: counting in systems with symmetries. 

In the previous paragraph, we computed the number of colorings of a network with distinguishable nodes. Counting colorings in networks with indistinguishable nodes is more complex as we have to deal with symmetries that may exist in the network structure. An efficient way to accommodate symmetries in the counting was discovered by Redfield and P\'{o}lya ~\cite{redfield1927theory,polya1937kombinatorische}. 

For a simple introduction to P\'{o}lya's counting theory, let's again consider the coloring of the linked pair (o-o) with white and grey. We start by following the approach from above and write the counting function as $G=(w+g)^2 = w^2 + 2wg + g^2$. But this counting function overcounts the possible colorings. Specifically, it counts the colorings in which there is one white and one grey node as two distinct configurations ($2wg$). However, since the nodes are indistinguishable, we want to count this case only as one configuration.

To correct for the double counting of the 1-white-1-grey configuration we might start by dividing the entire counting function by two, but this would result in an under-counting of those configurations where the symmetric nodes are colored with the same color. Hence we add an additional term ($X$) which will repair this under-counting. The result is 
\eq{
G_{\rm o-o}=\frac{(w+g)^2}{2} + X,
}
Now we only need to determine $X$. Because the first term now only counts the configurations where we have two white ($w^2$) nodes of two grey nodes ($g^2$) half, we need to add the two missing halves $X=(w^2+g^2)/2$. Hence our final counting function is  
\eq{
G_{\rm o-o}(w,g) = \frac{(w+g)^2}{2} + \frac{w^2+g^2}{2}
}
Evaluating for the number of colorings we now find 
\eq{
G_{\rm o-o}(1,1) = \frac{2^2}{2} + \frac{2}{2} = 3 
}
which is the correct result for coloring of the pair of indistinguishable nodes with two colors. 

If we instead consider the case of $C$ colors, we arrive at a counting function of the form 
\eq{
G_{\rm o-o} = \frac{(s_1)^2}{2} + \frac{s_2}{2}
}
where $s_1$ is the sum over all color variables, and $s_2$ is the sum over the squares of color variables, i.e.~all ways in which the two nodes can be colored identically. This form of the generating function is called the \emph{cycle index} and is often denoted $Z$, from the original German name \emph{Zyklenzeiger}. 
\subsection{P\'olya's Theorem} 
To formulate cycle indices for general problems it is useful to define 
\eq{
\label{eqsn}
s_n = \sum_{k=1}^C {c_k}^n.
}
where the $c_1 \ldots c_{C}$ are variables that are used to represent $C$ different colors. 

If we are only interested in the number of colorings we can then set all color variables to 1 to count the number of configurations, analogous to Eq.~(\ref{eqCompFirst}). As a result of this substitution the cycle variables become
\eq{
  s_n = \sum_{k=1}^C {1}^n = \sum_{k=1}^C = C.  
}

For example the number of ways in which we can color our indistinguishable linked pair, with $C$ colors is 
\eq{
G_{\rm o-o}(1,1,\ldots) = \frac{C^2}{2} +\frac{C}{2}.
}
So the number of colorings of the indistinguishable pair with 1000 colors is 500500.

To deal with more complex network structures it is useful to systematize the counting of symmetries: Mathematically speaking, a network symmetry can be conceptualized as an \emph{automorphism}, i.e.~a renumbering of nodes that leaves the adjacency matrix of the network unchanged. For example, a given automorphism may change the numbering of the nodes as follows 
\eq{
\label{eqExAutomorphism}
1\to 1, 2\to3, 3\to4,4\to2,
}
which means that the labels of nodes 2, 3, and 4 are cyclically permuted, while node 1 retains its label. The different sets of nodes that are exchanged with each other under the automorphism are called \emph{orbits}. In a more compact notation, we collect the labels of nodes that are exchanged in an orbit in braces in the respective order, and then list all the orbits. Hence, our example automorphism (Eq.~(\ref{eqExAutomorphism})) can be represented by (1)(234).

For the example of the linked pair there are two automorphisms: the trivial automorphism where each node retains it's label, and a nontrivial where the indices of the two nodes are swapped. Hence we can write the whole set of automorphisms, the \emph{permutation group}, as 
\eq{
S = \left\{(1)(2),(12) \right\}
}
where the first element $(1)(2)$ represents the trivial automorphism and the second element, $(12)$, represents the swapping of the indices. 

The number of elements of the permutation group $|S|$ is also the factor by which the node for distinguishable nodes might over-count colorings in the indistinguishable case. For the linked pair this gives us the factor $|S|=2$ that we know from above. In the following, all terms in the counting will be divided by this factor. 

We can now translate the automorphisms into terms of the counting function. We start with the trivial automorphism which exists in every network. In this automorphism, each node is in its own orbit. As a whole, the automorphism represents a contribution where each node is colored independently. Hence, we now replace each of the $N$ single-node orbits in the automorphism by the sum over colors $s_1$:
\eq{
(1)(2)(3)\ldots \to s_1 s_1 s_1 \ldots
}
After dividing by $|S|$ to correct for the over-counting configurations in which symmetric nodes are colored differently, we end up with the term ${s_1}^N/|S|$, where we are now under-counting the configurations where nodes in an orbit are colored identically. This is remedied by adding terms from the other automorphism. 

The non-trivial automorphisms may contain combinations of orbits of different sizes. To correct for the under-counting we need to consider terms in which each of these orbits is colored uniformly. Hence if the orbit contains two nodes we only want to consider the contributions from coloring both nodes identically. So, we can add two nodes of the first or two nodes of the second color or two nodes of the third color and so on. In terms of generating functions this choice is represented by the sum over squares of color variables $s_2$. In general, if an orbit contains $n$ nodes then the corresponding factor in the generating function is $s_n$, e.g:
\eq{
(12)(3)\ldots \to s_2s_1 \ldots
}

In summary P\'{o}lya's theorem implies that we can count the number of colorings in a network of indistinguishable nodes as follows:
\begin{enumerate}
\item Identify all automorphisms of the network and use them to write the symmetry group $S$. 
\item Turn $S$ into a cycle index $Z$: 
\begin{itemize}
  \item Replace all orbits in $S$ by factors $s_n$ where $n$ is the number of nodes in the orbit. 
  \item The cycle index $Z$ is the sum over all automorphisms, divided by the size of the symmetry group $|S|$.
\end{itemize}
\item  To count the number of colorings from the cycle index, set all $s_n$ to the number of colors, $s_n=C$.  
\end{enumerate}

 \begin{figure}[ht]
    \centering
    \includegraphics[width=10cm]{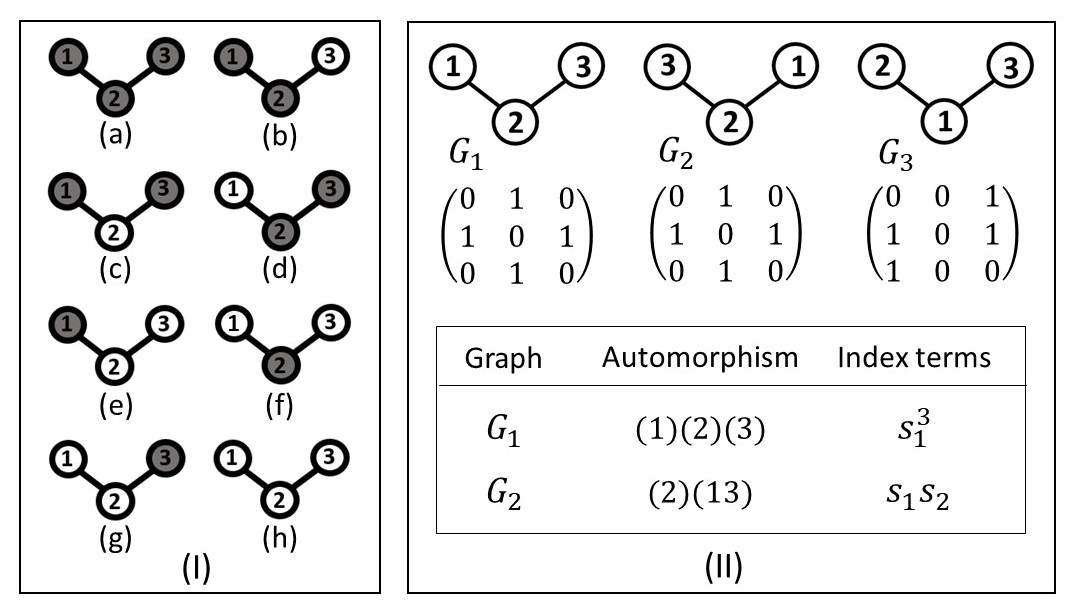}
    \caption{Coloring a three-node chain. There are 8 distinguishable colorings, but only 6 if nodes are indistinguishable (I). 
    The labeled graph ($G_1$ in II) has an adjacency matrix that is identical to a graph where the numbers on the end-nodes of the chain are exchanged ($G_2$ in II), which reveals that the exchanging of the end nodes is a nontrivial automorphism. However, not every renumbering of the nodes defines an automorphism as the example of $G_3$ shows. The nontrivial automorphism, represented by (2)(13) leads to the term $s_1s_2$ in the counting function, whereas the trivial automorphism, (1)(2)(3), yields a term ${s_1}^3$}.
    \label{fig:ColoringEx}
\end{figure}  

Another example of this procedure is shown in Fig.~(\ref{fig:ColoringEx}). Here we consider a three-node chain (o-o-o). In addition to the trivial automorphism, the chain has a flip symmetry where the first and the third node are exchanged. The figure illustrates that this flip indeed leaves the adjacency matrix unchanged, whereas for instance  exchanging the labels of nodes 1 and 2 leads to a different adjacency matrix. 

The symmetry group of the three-node chain is
\eq{
S = \left\{ (1)(2)(3), (13)(2) \right\}
}
To find the cycle index we replace the orbits with the color sums $s_n$, which yields ${s_1}^3$ for the trivial automorphism ((1)(2)(3)), and $s_2s_1$ for the flip symmetry ((13)(2)). Summing over these terms and dividing by the size of the symmetry group $|S|=2$ we arrive at the cycle index 
\eq{Z_{o-o-o} = \frac{{s_1}^3+s_2s_1}{2}}
If we are interested in the number of ways in which the three-node chain can be colored with two colors, we can then set $s_n=2$ for all $n$ which yields the correct answer, $6$. 

\subsection{Counting Trees} 
\label{subsec:Counting-Trees} 
The final ingredient that we need to count chemical isomers is Read's insight that the generating functions can be recursively defined~\cite{read}. To illustrate this idea, let's count the number of rooted trees consisting of nodes of degrees 1 and 3 and starting from a root of degree 2. 

\begin{figure}[ht]
    \centering
    \includegraphics[width=10cm]{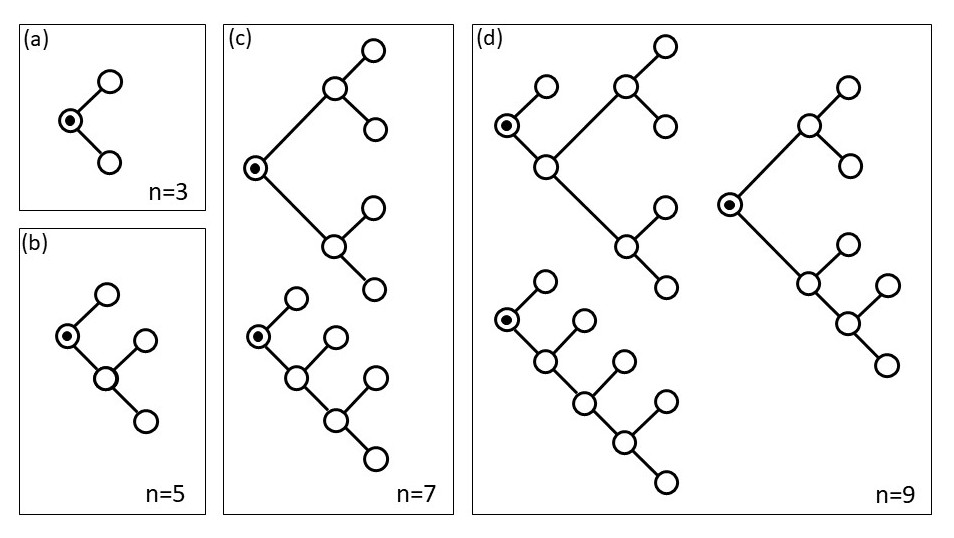}
    \caption{Panels (a)-(d) illustrate all possible unlabeled trees of sizes 3,5,7, and 9, where each node has either degree one or degree three. For example, factor 2 in front of $x^7$ in Eq.~(\ref{eqTree}) shows that there are two distinct tree structures with 7 nodes (panel(c)). Note that the circle with a black dot at its center in the panels represents the trees' root. }
    \label{fig:Tree}
\end{figure}  

In the generating function $T$ that we are about to derive we represent the nodes by $x$. If multiplied out to polynomial form the generating function will then contain one term $x^n$ for each possible tree that contains $n$ nodes. Hence, it will start with 
\eq{ \label{eqTree}
T=x^3+x^5+2x^7+\ldots
}
as there is only one possible tree with three nodes, one possible tree with 5 nodes, but two distinct trees with 7 nodes (Fig.~(\ref{fig:Tree})).

While we might thus identify the first terms of $T$ using our intuition, it is clear that there is an infinity of further terms as the possible size of the trees is not bounded. To derive the generating function nevertheless, let's first derive the generating function, $B$, for a branch of the network that we enter via a link. 

If we follow a link we can either reach a node of degree 1 or a node of degree 3. In our generating function a node of degree 1 the whole branch consists only of a single node, so $x$. If we reach a node of degree 3, the branch consists of this node (again represented by $x$) plus all the nodes that we find in the two sub-branches starting in this node. In terms of generating functions the `plus´ in the previous sentence amounts to a multiplication of terms, so we might represent the case where we reach a node of degree 3 by $xY$, where $Y$ is a generating function that counts the ways in which two sub-branches may add up to a given number of nodes. 

Because entering a branch via a link may lead us either to a node of degree 1 or a node of degree 3 the generating function for the branch has the form 
\eq{
\label{eqBstart}
B=x+xY
}
where the first term represents the case where we arrive at a node of degree one and the second term represents the case where we arrive at a node of degree three. 

To determine the factor $Y$ let's first recall that $Y$ should be a generating function that has a term $x^n$ for each way in which we can attach two subtrees that have $n$ nodes in total. Let's practice this in some simplified cases: For example, if we only had subtrees of size one, the only option would be to attach two subtrees that have two nodes in total and hence $Y=x^2$. 

Now consider the case where we have the choice between two types of subtrees, a small one, with one node $a=x$, or a slightly bigger one, with three nodes $b=x^3$. In this case, we could attach two small subtrees, two larger subtrees, or one large and one small subtree, and hence $Y=a^2+ab+b^2$. This expression looks exactly as if we are coloring the two links of our focal node with two colors `a' and `b'.
Indeed, we can use the same reasoning as before to write 
\eq{ 
Y=\frac{{s_1}^2+s_2}{2},
}
as a result of the mirror symmetry between the two links from the focal node. The beauty of this equation is that it holds independently of the list of subtrees that we have at our disposal. In the case where we only have the small subtree $a$, this $a$ is our only `color',
hence $s_1=a=x$ and $s_2=a^2=x^2$ (compare Eq.~(\ref{eqsn}) with $c_1=a$). In the case where we have both $a$ and $b$, we have $s_1=a+b$, $s_2=a^2+b^2$.
 
Of course in the case in which we are really interested we have not only one or two different subtrees at our disposal, but the set of all possible subtrees. In this case $s_1$ becomes a sum over all possible subtrees, where each term $x^n$ again represents a subtree with $n$ nodes, we have already given this sum a name, it is  $B(x)$, and hence $s_1=B(x)$. Similarly, $s_2$ is the sum over all possible subtrees where the term representing each subtree is squared. Because the terms are monomials we can write this sum as $s_2=B(x^2)$. Hence we can write
\eq{
Y=\frac{B^2(x) + B(x^2)}{2}.
}
Substituting back into Eq.~(\ref{eqBstart}) we arrive at 
\eq{
\label{eqBranchResult} 
B(x)=x\left(1+\frac{B^2(x) + B(x^2)}{2}\right)
}
which gives us an implicit equation that describes the set of potential branch structures.

To find a generating function that counts the whole set of trees we still need to deal with the degree-two node at the root. But this is easy now: The root is just another node ($x$) at which two branches start ($Y$), and hence it is represented by
\eq{
\label{eqTreeGF}
G(x)=xY=x\frac{B^2(x) + B(x^2)}{2}.
}

\subsection{Result Generation}
\label{subsec:Result-Gen} 
%
In some cases we can just solve the set of generating functions to find explicit forms. But even if this is not possible, it is often possible to extract the information that we seek. 

If we are interested in the full set of counts we could Taylor expand $B$ and $G$, possibly using a software tool for symbolic mathematics. However, a more efficient numerical solution is typically to partially undo the generating functions by recalling their construction as 
\eq{
\label{eqTreeHalfway}
B(x)=\sum_k b_k x^k, \quad\quad\quad G(x)=\sum_k g_k x^k. 
}
substituting into Eq.~(\ref{eqBranchResult}) yields,
\eq{
\label{eqIThalfway}
\sum b_k x^k=x\left(1+\frac{\left(\sum b_k x^k\right)^2 + \sum b_k x^{2k}}{2}\right).
}
To simplify again we can now match the coefficients in front of the factor $x^k$ in this equation. On the left the factor in front of the $x^k$ is simply $b_k$, but on the right there are three separate terms:
The first term is simply an $x$ which we only have to take into account when the $k$ we are interested in is $k=1$, hence in an equation for the factors in front of the $b^k$ terms, this term is represented by the Kronecker delta, $\delta_{k1}$.

The second term is $x(\sum b_l x^l)^2/2$ where we have changed the summation index from $k$ to $l$. We now ask which prefactor will appear in front of the term where the square over the sum results in a factor $x^k$. We can write 
\eq{
\label{eqSquareDecomp}
\left(\sum_l b_l x^l\right)^2 = \left(\sum_i b_i x^i\right)\left(\sum_j b_j x^j\right) = \sum_{ij} b_i b_j x^{i+j}   
}
Because the corresponding term in Eq.~(\ref{eqIThalfway}) is multiplied by an additional $x$ we are looking for the terms in Eq.~(\ref{eqSquareDecomp}) that contain $x^{k-1}$ hence we consider only those terms of the sum where $i=k-1-j$ which eliminates the summation over $i$. Taking the additional division by 2 in Eq.~(\ref{eqIThalfway}) into account we arrive at the factor $\sum b_{k-1-j}b_j/2$.

The final term in Eq.~(\ref{eqIThalfway}) is $x\sum_l b_l x^{2l}/2 $, where we have again changed the index of summation to $l$. We find a factor $x^k$ for $k=2l+1$ or conversely $l=(k-1)/2$. Hence the prefactor in front of the $x^k$ term is $b_{(k-1)/2}/2$ if $k$ is odd or zero otherwise. 

Putting all three contributions together, we find that Eq.~(\ref{eqIThalfway}) implies
\eq{
b_k = \delta_{1k} +  \frac{\left(\sum b_{k-1-j}b_j\right)+b_{(k-1)/2}}{2}
}
where the $b_{(k-1)/2}$ is zero for even values of $k$. This equation gives us an efficient way to find the $b_k$ by iteration. For example 
\eq{
\begin{array}{r c c c l}
b_0 &=&  0 + (0+0)/2 &=& 0 \\
b_1 &=&  1 + ({b_0}^2+b_0) /2  &=& 1 \\
b_2 &=&  0 + (2b_0b_1+0)/2  &=& 0 \\
b_3 &=&  0 + (2b_0b_2+{b_1}^2+b_1)/2 &=& 1 \\
b_4 &=&  0 + (2b_0b_3+2b_1b_2+0)/2 &=& 0 \\
b_5 &=&  0 + (2b_4b_0+2b_1b_3+{b_2}^2+b_2)/2 &=& 1
\end{array}
}
Similarly identifying the factors in front of $x^k$ term in Eq.~(\ref{eqTreeHalfway}) leads to
\eq{
g_k = \frac{\left(\sum b_{k-1-j}b_j\right)+b_{(k-1)/2}}{2} 
}
which yields 
\eq{
g_0 = 0,\quad g_1=0, \quad g_2=0, \quad g_3=1, \quad g_4=0, \quad g_5=1, \ldots
}
This result shows that under the rules described above there is no way to make trees with 0, 1, 2, or 4 nodes, but exactly 1 way to make a tree with 3 or 5 nodes. 


\section{Counting Branching Isomers\label{sec:GCF}} 
%

We now use the approach described above to count the number of isomers of certain classes of chemical molecules. In particular, we are interested in organic molecules in which treelike branches of atoms are attached to a more complex, and typically cyclic structure acting as a root.

\begin{figure}[h]
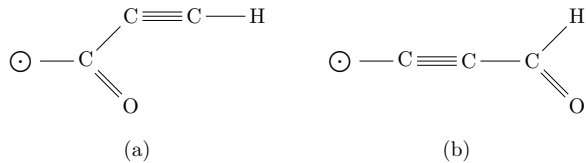
 
\centering
\scalebox{0.8}{\chemname{\chemfig{\bigodot-C(-[:45]C~C-H)(=[:-45]O)}}{(a)}}
\hspace{.6cm}
\scalebox{0.8}{\chemname{\chemfig{\bigodot-C~C-C(-[:45]H)(=[:-45]O)}}{(b)}}
\caption{The representation of the two distinct molecular configurations formed by one hydrogen, one oxygen, and three carbon atoms($-C_{3}OH$). The structural differences arise due to variations in bonding patterns. Note that the circle with a black dot at its center represents the root of the molecules.}
\label{fig:DiffConfig}
\end{figure}

In this section, we identify the counting function for branches and generate counts for different branch structures starting from a single link (a monovalent bond). Going beyond previous mathematical work~\cite{faulon2005enumerating}, we also consider the role of oxygen and allow multiple bonds between carbon atoms in the branches, which is appropriate in the context of the application (Fig.~(\ref{fig:DiffConfig})). 

A consequence of allowing double bonds and oxygen is that there can be different isomers with the same number of carbon atoms but different numbers of hydrogen or oxygen atoms. In contrast to many previous works, it is therefore no longer sufficient to use univariate generating functions that only account for the carbon skeleton of a molecule. Instead, we use multivariate generating functions that account for carbon, oxygen, and hydrogen.


\subsection{Generating Function for Chemical Molecules}
\label{subsec:Model-molecules} 
%
%
We denote this generating function for the number of trees that start from a root via a single bond as $A(c,o,h)$, where $c$, $o$, and $h$ represent carbon, oxygen, and hydrogen atoms, respectively. We proceed by considering the different types of atoms that can connect to the bond at the root of the branch (Fig.~(\ref{fig: Counting function A, B, T})).

If the bond connects to a single hydrogen atom, then this immediately ends the tree, and the corresponding counting function is just $A_0=h$, where $h$ is a placeholder for the hydrogen atom. For our application to the dissolved organic matter in the oceans, it would not be useful to count structures that contain an oxygen atom as a bridge between carbon atoms in a branch structure, as this would be very unstable. Therefore, the only other atom that is allowed to follow the oxygen is a hydrogen atom that is attached via a single link, forming an $oh$ group. In this case, the term for the counting function is $A_1=oh$. 

\begin{figure}[ht]
    \centering
    \includegraphics[width=12cm]{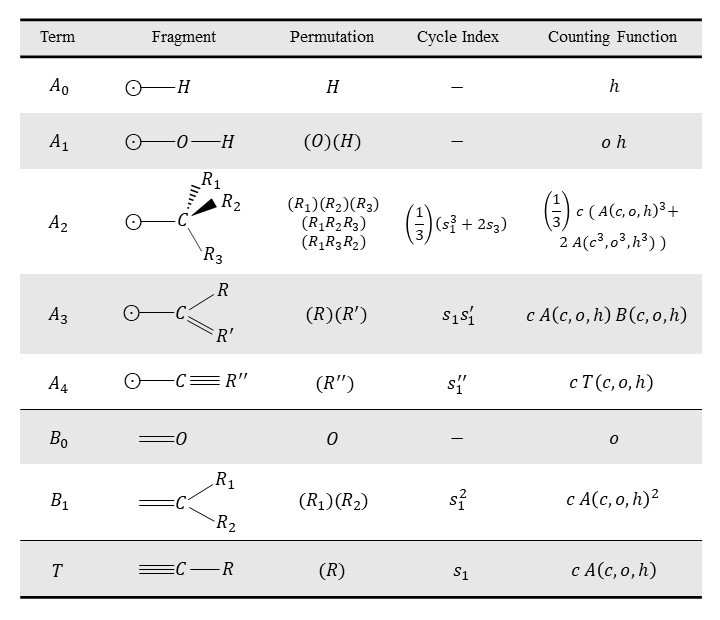}
    \caption{ Possible chemical fragments attached to a rooted single bond, a double, and triple bond  with the corresponding cycle index and contribution in the generating counting function $A$. Trees $R$, $R'$, and $R^"$ connect to single, double, and triple bonds, respectively.}
    \label{fig: Counting function A, B, T}
\end{figure}

The root bond can also connect to a carbon atom, and, in this case, it is useful to distinguish some subcases: a) The root link could lead to a carbon atom, where the three remaining links are the start of three individual subbranches. b) The carbon atom could be the start of two independent subbranches, one starting with a double bond, and one starting with a single bond. c) The carbon atom could lead to a single subbranch starting with a triple bond. 

The first case gives us three new bonds to which trees starting from a single bond can be attached. We already defined the counting function for a tree starting from a single bond as $A$. But now that we have three symmetric single bonds, we need to take the cycle index of the symmetry into account, which is $(s_1^3 + 2 s_3)/3$, and hence the counting function for this case is $A_2=c( A(c,o,h)^3 + 2A(c^3,o^3,h^3))/3$, in which, the multiplier $c$ accounts for the carbon atom to which the outgoing trees are attached. 

In the second case, there is no symmetry to consider because the outgoing single and double bonds are distinguishable from each other. If we define a counting function $B$ for trees that start with a double bond, then the counting function for this case will be the multiplication of two functions, i.e.~$A_3=cA(c,o,h)B(c,o,h)$. 

Finally, the counting function for the third case is $A_4= cT(c,o,h)$, where the function $T$ is defined as a counting function for trees that start with a triple bond. Given the construction rules, only one carbon atom with an outgoing single bond can join this triple bond, i.e.~$T(c,o,h)= cA(c,o,h)$. Therefore, we obtain $A_4=c^2A(c,o,h)$ for the last case contribution.

Ultimately, as detailed in Sec.~\ref{sec:Counting-Theory} , the generating function of our rooted chemical molecule is equal to the sum of the counting functions of all possible chemical fragments $A_i$, which leads to
\begin{multline} 
A(c,o,h)=h+oh+ \frac{1}{3} c \ ( A(c,o,h)^3 + 2A(c^3,o^3,h^3)) \\  +c \ A(c,o,h)B(c,o,h)+ c^2 \ A(c,o,h). 
\label{eq:A(c,o,h)}
\end{multline}   

At this point, it is still necessary to determine the counting function $B$.
By considering the different types of atoms that can connect to the double bond, we can identify different subbranches (Fig.~(\ref{fig: Counting function A, B, T})). 
The first case might be a single oxygen atom linked to the double bond, and thus the corresponding counting function is just $B_0=o$.  

The double bond can also connect to a carbon atom, leading to two different subcases: a) The carbon atom with two outgoing single bonds or b) The carbon atom with one outgoing double bond. 

The double bond can also connect to a carbon atom that has two outgoing single bonds. Each of these single bonds can be attached to a tree represented by the generating function $A$. In this case, the resulting counting function is the product of the two generating functions $A$ 
yielding $B_1= cA(c,o,h)^2$.It is important to note that we do not consider symmetry at this stage, as we are enumerating stereoisomers. 
Finally, the total counting function $B$ for a tree that begins with a double bond is

\begin{equation} 
B(c,o,h)=o+  c A(c,o,h)^2 . 
\label{eq:B(c,o,h)}
\end{equation}
The last two equations (Eq.~(\ref{eq:A(c,o,h)}) and~(\ref{eq:B(c,o,h)})) enable us to calculate the generating functions recursively up to any given power of $c$, $o$, and, $h$. 


\subsection{Deriving Recursive Relations}
\label{subsec:Recursive-Rel} 
%
%
In the previous section we arrived at functional equations for the generating functions $A$ and $B$. As in the introductory example from Secs.~\ref{subsec:Counting-Trees} and~\ref{subsec:Result-Gen}, we now partially undo the generating functions to find iteration rules that allow for an efficient computation of the desired numbers of isomers.

We start by recalling that generating functions such as $A$ can be represented by multivariate powerseries, i.e. 
 \begin{eqnarray}
A(c,o,h)=\sum_{l=0} \sum_{m=0} \sum_{n=0} a_{l,m,n}\; c^lo^mh^n,  
\label{eq:powerseries-A}
\end{eqnarray} 
where $a_{l,m,n}$ are coefficients representing the number of isomers with $l$ carbon, $m$ oxygen, and $n$ hydrogen atoms. 

We define a similar power series for 
\begin{eqnarray}
B(c,o,h)=\sum_{l=0} \sum_{m=0} \sum_{n=0} b_{l,m,n}\; c^lo^mh^n,  
\label{eq:powerseries-B}
\end{eqnarray}  
where $b_{l,m,n}$ are coefficients corresponding to the fragments connected to a double bond in the acyclic organic molecule. 

To establish a relation among coefficients $a_{l,m,n}$ and $b_{l,m,n}$, we replace every occurrence of $A(c,o,h)$ and $B(c,o,h)$ in Eq.~(\ref{eq:A(c,o,h)}) and Eq.~(\ref{eq:B(c,o,h)}) by the power series in Eq.~(\ref{eq:powerseries-A}) and Eq.~(\ref{eq:powerseries-B}). For $A$ this yields
\begin{equation}
\begin{array}{r c l}
\sum_{l,m,n} a_{l,m,n} c^lo^mh^n &=& h +oh
 + c^2\sum_{l,m,n} a_{l,m,n} c^{l}o^{m}h^{n} \\ 
 &+& \frac{2}{3}c\sum_{l,m,n} a_{l,m,n} c^{3l}o^{3m}h^{3n} \\
 &+&  \frac{1}{3}c\left (\sum_{l,m,n} a_{l,m,n} c^lo^mh^n\right )^3 \\ 
&+& c\left(\sum_{l,m,n} a_{l,m,n} c^lo^mh^n\right) \left (\sum_{l,m,n} b_{l,m,n} c^lo^mh^n \right)\label{eq:equal-to-left}. 
 \end{array}
\end{equation}

We can now extract the coefficients $a_{l,m,n}$ by matching the powers of $c$, $o$, and $h$. As the first two terms on the right $h$ and $oh$ are the only ones that do not contain a $c$ we can see that all $a_{0,l,m}=0$ except for $a_{0,0,1}=1$ and $a_{0,1,1}=1$. 

To find an iteration rule for the remaining terms, we follow the approach from Secs.~\ref{subsec:Counting-Trees} and~\ref{subsec:Result-Gen}. Because a lot of information is now contained in the indices we introduce the notation $a_{l,m,n}=[l,m,n]$ and $b_{l,m,m}=\{l,m,n\}$ which allows us to write,

  \begin{align}
[l, m, n] &= \delta_{l,0}\delta_{m,0}\delta_{n,1}
           + \delta_{l,0}\delta_{m,1}\delta_{n,1} 
           + [l-2, m, n] 
           + \tfrac{2}{3}[{\tfrac{l-1}{3}}, {\tfrac{m}{3}}, {\tfrac{n}{3}}] \notag \\
           &\phantom{{}=} + \tfrac{1}{3} \sum_{i,j,k} \sum_{x,y,z}
           [l-i-x-1, m-j-y, n-k-z][i, j, k][x, y, z] \notag \\
           &\phantom{{}=} + \sum_{i,j,k}[l-i-1, m-j, n-k]\{i, j, k\}
           \label{eq:almn}
           \\
\{l, m, n\} &= \delta_{l,0}\delta_{m,1}\delta_{n,0}
           + \sum_{i,j,k}[l-i-1, m-j, n-k][i, j, k].
           \label{eq:blmn}
\end{align}

where $\delta$ is the Kronecker delta operator, and terms with fractional coefficients negative or 
fractional parameters are considered to be zero, e.g.~$[-1,1,1]=[\frac{1}{3},\frac{4}{3},\frac{4}{3}]=0$.

Recurrence Eqs.~\ref{eq:almn} and~\ref{eq:blmn} constitute our main results. Together with the initial conditions $a_{0,0,1}=1 $, $ a_{0,1,1}=1$, and $b_{0,1,0}=1 $, these recurrence relations enable the numerical computation of the entire sequence of isomers.
 

\begin{table}
    \centering
    \begin{tabular}{l | c  |l| c |l| c }
    $l,m,n$ & $a_{l,m,n}$& $l,m,n $& $ a_{l,m,n} $ & $ l,m,n $& $ a_{l,m,n} $ \\\hline 
    $1,0,3$ &  $1$  &     $6,4,1$ &  $5$ &      $17,4,13$  &  $171786815460$  \\
    $1,1,1$ &  $1$  &     $6,4,3$ &  $377$&     $17,4,15$  &  $744906775092$  \\
    $1,1,3$ &  $1$  &     $6,4,5$ &  $4065$&     $17,4,17$  &  $2359219102802$ \\
    $1,2,1$ &  $1$  &     $6,4,7$ &  $15466$&      $17,4,19$  &  $5565012462786$  \\
    $1,2,3$ &  $1$  &     $6,4,9$ &  $24554$&      $17,4,21$  &  $9854769878432$  \\
    $1,3,3$ &  $1$  &     $6,4,11$&  $16480$&     $17,4,23$  &  $13073894296000$  \\
    $2,0,1$ &  $1$  &     $6,4,13$&  $3814$ &     $17,4,25$  &  $12838008640602$   \\
    $2,0,3$ &  $1$  &     $6,5,1$ &  $5$ &     $17,4,27$  &  $9114910025258$   \\
    $2,0,5$ &  $1$  &     $6,5,3$ &  $325$ &     $17,4,29$  &  $4504585195716$   \\
    $2,1,1$ &  $1$  &     $6,5,5$ &  $4521$ &     $17,4,31$  &  $1456704376932$   \\
    $2,1,3$ &  $5$  &     $6,5,7$ &  $19242$ &     $17,4,33$  &  $275056197386$   \\
    $2,1,5$ &  $3$  &     $6,5,9$ &  $32840$ &     $17,4,35$  &  $22861825074$  \\
    \vdots  & \vdots &    \vdots     &  \vdots &  \vdots &  \vdots  \\
    $3,3,7$ &  $22$ &       $11,5,7$ &   $6981764$ &   $20,5,9$ &  $74987043740$ \\
    $3,4,1$ &  $1$ &     $11,5,9$ &   $49479850$ &  $20,5,11$ &  $ 1030061453840$\\
    $3,4,3$ &  $14$ &    $11,5,11$ &  $200330066$ & $ 20,5,13$ &  $9328016200076$\\
    $3,4,5$ &  $40$ &   $11,5,13$ &  $497595694$ & $ 20,5,15$ & $ 59479130926080$\\
    $3,4,7$ &  $22$ &   $11,5,15$ &  $776011070$ & $ 20,5,17$ &  $278822367402127$ \\
    $3,5,3$&  $4$&   $11,5,17$ &  $751059478$ & $ 20,5,19$ &  $988495307859024$   \\
    $3,5,5$&  $21$ &   $11,5,19$ &  $431818480$ & $ 20,5,21$  &  $2698106174051362$\\
    \vdots     & \vdots  &     \vdots     &  \vdots &  \vdots  &  \vdots\\
   
    \end{tabular}
    \vspace{1cm} 
    \caption{Coefficients $a_{l,m,n}$ are the number of isomers for a rooted general acyclic compound with $l,m,$ and $n$ carbon, oxygen, and hydrogen atoms.}
    \label{tab:listIsomers}
\end{table}

We have determined and listed the number of isomers for a given sum formula with $l,m,$ and $n$ carbon, oxygen, and hydrogen atoms in Tab.~(\ref{tab:listIsomers}). Indices $l,m,$ and $n$  are the exponents of placeholders $c$, $o$, and $h$ in counting series, and the numbers in the second columns are their coefficients; for instance, the rooted sum formula $C_6O_4H_5$ could have 4065 different molecular configurations. 

Note that the numbers in the table indicate the number of both structural and stereo isomers; however, one can choose to count only structural isomers by adjusting the cycle indeces in terms $A_i$ and $B_i$ (Fig.~(\ref{fig: Counting function A, B, T}) ). For instance, to only enumerate the structural isomers, the permutations $(R_1)(R_2R_3)$, $(R_3)(R_1R_2)$, and $(R_2)(R_3R_1)$ must be added to the fragment $A_2$, which means including the term $3s_1s_2$ in the cycle index. We also need to take the symmetry of the chemical fragment $B_2$ into account and would get $(s_{1}^2 +s_{2})/2$ as the cycle index for term $B_2$.

\begin{figure}[ht]
   \centering
    \includegraphics[width=12cm]{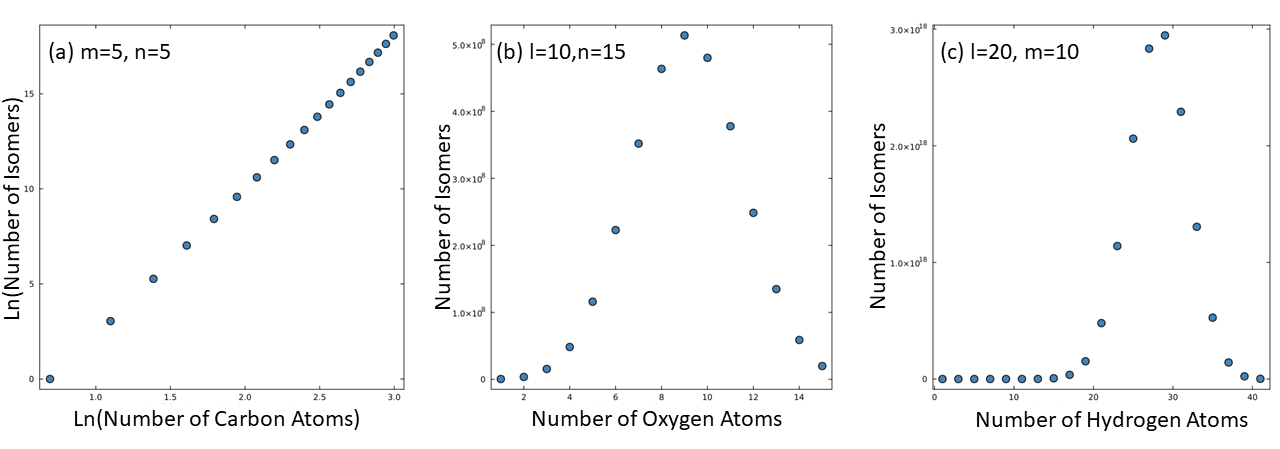}
    
    \caption{Panels (a), (b), and (c) illustrate the growth of isomers with respect to the number of carbon (constant $m$ and $n$), oxygen (constant $l$ and $n$), and hydrogen (constant $l$ and $m$) atoms, respectively. Note that both axes in figure (a) are plotted on a natural logarithmic scale.}
    \label{fig:IsoPlot}
\end{figure} 

As can be seen in Fig.~(\ref{fig:IsoPlot}), the number of isomers grows explosively with the number of atoms, due to the enormous number of possible combinatorial arrangements. 

Carbon atoms, through the formation of multiple covalent bonds, drive molecular branching and expansion, thereby generating an enormous increase in the diversity of structural configurations. In contrast, oxygen and hydrogen atoms terminate the molecular tree branches. As the number of oxygen and hydrogen atoms initially increases, the number of isomers rises sharply, reaches a peak, and then declines. The reason for this decline at higher oxygen and hydrogen counts is that, for a fixed number of carbon atoms, due to bonding constraints, the number of remaining positions available for oxygen and hydrogen atoms becomes more limited.


\section{Isomers for ring structures\label{sec:RingStructures}} 
%
%
As an application we can now use the coefficients $a_{l,m,n}$ in Eq.~(\ref{eq:almn}) to compute the number of isomers for several types of cyclic compounds, such as monocyclic aromatics (molecules with one ring, Fig.~\ref{fig: aromatic examples}(a) and polycyclic aromatics (molecules with two stuck rings, Fig.~\ref{fig: aromatic examples}(b)). 

\subsection{Isomers for Benzene Skeleton\label{subsecBenzene}}
\begin{figure}[h]
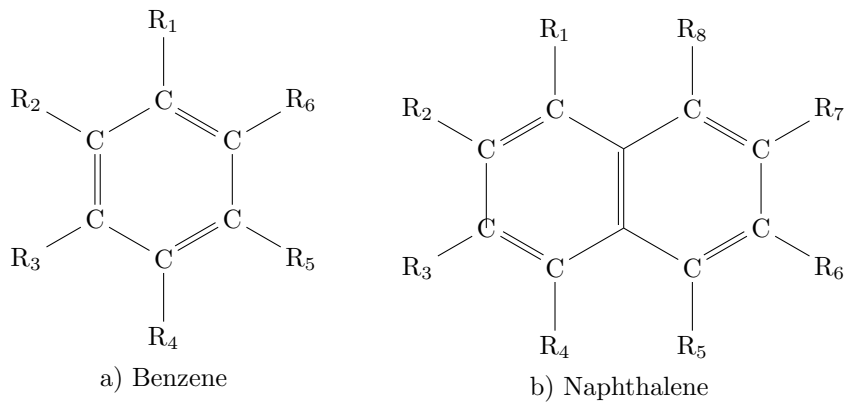
 
\centering
\chemname{\chemfig{C*6((-R_{3})-C(-R_{4})=C(-R_{5})-C(-R_{6})=C(-R_{1})-C(-R_{2})=)}}{a) Benzene} 
\hspace{.9cm}
\chemname{\chemfig{*6(C(-R_{3})=C(-R_{4})-(*6(-C(-R_{5})=C(-R_{6})-C(-R_{7})=C(-R_{8})-))=-C(-R_{1})=C(-R_{2})-)}}{b) Naphthalene } 
\caption{Two cyclic molecular structures. The acyclic molecules $R_i$  are attached to each corner, where the carbon atoms at the corners of the rings are now the roots of these chains.  a) A monocyclic aromatic molecule. The permutation $(R_3)(R_6)(R_1R_5)(R_2R_4)$, for instance, on (a) with cycle index $s_{1}^{2}s_{2}^{2}$ leaves the benzene unchanged. b) A polycyclic aromatic. The permutation $(R_1R_4)(R_2R_3)(R_5R_8)(R_6R_7)$, for instance, on (b) with cycle index $s_{2}^{4}$ preserves the adjacency matrix of naphthalene. }
\label{fig: aromatic examples}
\end{figure}

To construct the most general organic molecule rooted in a ring, we attach chains $R_i$ to the carbon atoms at the corners of the hexagon (Fig.~\ref{fig: aromatic examples}(a) ). These chains $R_i$ are the same acyclic organic molecules formed by the combinatorial arrangements of chemical fragments shown in Fig.~(\ref{fig: Counting function A, B, T}) in Sec.~\ref{sec:GCF}. 

The cycle indices for each symmetry operation that may be executed on the benzene and preserve the adjacency matrix can be found in~\cite{faulon2005enumerating}. The cycle index of the automorphism group $Z$ for the benzene ring is


\begin{eqnarray}
Z=\frac{1}{12}(s_{1}^{6}+2s_{6}+2s_{3}^{2}+4s_{2}^{3}+3s_{1}^{2}s_{2}^{2}). 
\label{eq:CycleRing}
\end{eqnarray} 

We define a counting function $D(c,o,h)$ for the benzene ring as the power series 
\begin{eqnarray}
D(c,o,h)=\sum_{l=0} \sum_{m=0} \sum_{n=0} d_{l,m,n} \; c^lo^mh^n
\label{eq:PowerRing}
\end{eqnarray} 
in which coefficients $d_{l,m,n}$ encode the number of isomers. We can write the counting function explicitly by substituting our previously dervived counting function for trees $A(c^n,o^n,h^n)$ for instances of the cycle index $s_n$ in the counting function of the ring 
\begin{multline}
D(c,o,h)= \frac{1}{12} \;c^6  \Bigl( A(c,o,h)^{6}+2A(c^6,o^6,h^6)+2A(c^3,o^3,h^3)^{2}\\ +  4A(c^2,o^2,h^2)^{3}+3A(c,o,h)^{2}A(c^2,o^2,h^2)^{2} \; \Bigl ).
\label{eq:GFRing}
\end{multline}
where the factor $c^6$ counts the six carbon atoms of the ring. 

Substituting Eq.~(\ref{eq:PowerRing}), and Eq.~(\ref{eq:powerseries-A}) in Eq.~(\ref{eq:GFRing}), and conducting similar calculations as in Sec.~\ref{sec:Counting-Theory}, the counting coefficients $d_{l,m,n}$ are connected to the counting coefficients $a_{l,m,n}$ recursively by the following equation  
\begin{eqnarray} 
d_ { l,m,n} = \frac{1}{12} ( \ D_{1}+ D_{2}+ D_{3}+ D_{4}+ D_{5} \ ),
\end{eqnarray}  
where
\begin{eqnarray} 
D_1  &=&  \sum_{w,w',w"}\sum_{v,v',v''} 
\sum_{u,u',u''}\sum_{t,t',t''}  \sum_{s,s',s''}[s,s',s''][t,t',t''][u,u',u''] \\ 
 & & [v,v',v''][w,w',w''] [x,x',x''] ,  \\
D_2 &=&  2 \ [\frac{l-6}{6},\frac{m}{6} ,\frac{n}{6} ]  ,   \\
D_3 &=&   2 \sum_{s,s',s'\!'}   [\frac{l-3s-6}{3},\frac{m-3s'}{3} ,\frac{n-3s''}{3}] [s,s',s'']  \\
D_4 &=& 4 \sum_{s,s',s''} \sum_{t,t',t''}  [s,s',s''] [t,t',t''] \\
& & [\frac{l-2s-2t-6}{2},\frac{m-2s'-2t'}{2} ,\frac{n-2s''-2t''}{2}]  \\
D_5 &=& 3\sum_{s,s',s''}  \sum_{t,t',t''}  \sum_{u,u',u''} 
[y,y',y''][u,u',u''] [t,t',t''] [s,s',s''] .
\end{eqnarray}
where indices in coefficient $a_{x\!,x'\!,x'\!'} $ are 
\begin{eqnarray}
x  &=&  l-s-t-u-v-w-6,  \\
x' &=&  m-s'-t'-u'-v'-w',  \\
x'' &=&  n-s''-t'' -u''-v''-w'', 
\end{eqnarray}
and indices in coefficient $a_{y\!,y'\!,y'\!'}$ are
\begin{eqnarray}
y  &=&  l-s-2t-2u-6 ,\\
y' &=&  m-s'-2t'-2u',\\
y'' &=&  n-s''-2t''-2u''. 
\end{eqnarray}

The resulting coefficients, i.e., the number of isomers for benzene rings for several initial numbers of carbon, oxygen, and hydrogen, are listed in Tab.~(\ref{tab:listIsomersRing}).

\subsection{Isomers for Naphthalene Skeleton\label{subsecnaphthalene}}
Similar to  the previous case, a naphthalene ring is now the root to whose corners the acyclic organic molecules $R_i$ are connected. The cycle index and, therefore, the counting generating function are
\begin{eqnarray}
Z=\frac{1}{4}(s_{1}^{8}+3s_{2}^{4}),  
\label{eq:CycleNaphthalen}
\end{eqnarray} 

\begin{eqnarray}
F(c,o,h)= \frac{1}{4}c^{10}  \Bigl( A(c,o,h)^{8}+3A(c^2,o^2,h^2)^{4} \Bigl  ),
\label{eq:GFNaphthalen}
\end{eqnarray} 
 where now 
 \begin{eqnarray}
 F(c,o,h)= \sum_{l=0} \sum_{m=0} \sum_{n=0} f_{l,m,n} \; c^lo^mh^n 
 \label{eq:PowerNaphthalene}
 \end{eqnarray} 
is the power series of the naphthalene ring. 

By substituting the power series expressions from Eq.~(\ref{eq:PowerNaphthalene}) and Eq.~(\ref{eq:powerseries-A}) into Eq.~(\ref{eq:GFNaphthalen}), and applying analogous computations as those detailed in Sec.~\ref{sec:Counting-Theory}, we derive the following recursive relation that links the counting coefficients $f_{l,m,n}$ to $a_{l,m,n}$ as
\begin{multline} 
f_{l,m,n} = \frac{1}{4} (\sum_{s,s',s'\!'}\sum_{t,t',t'\!'}\sum_{u,u',u'\!'}\sum_{v,v',v'\!'}
\sum_{w,w',w'\!'}\sum_{x,x',x'\!'}\sum_{y,y',y'\!'}  \ a_{y,y',y''} \\
a_{x\!,x',x'\!'} \   a_{w\!,w'\!,w'\!'} \   a_{v\!,v'\!,v'\!'} \ a_{u\!,u'\!,u'\!'} 
\  a_{t\!,t'\!,t'\!'}  
\  a_{s\!,s'\!,s'\!'}  \ a_{z\!,z'\!,z'\!'} \ \ +  \\ 3 \sum_{s,s',s'\!'}\sum_{t,t',t'\!'}\sum_{u,u',u'\!'}   a_{u\!,u',u'\!'} \ a_{t\!,t'\!,t'\!'} \  a_{s\!,s'\!,s'\!'} \ a_{r\!,r'\!,r'\!'} ), 
\end{multline}  
where indices in coefficient $a_{z\!,z'\!,z'\!'} $ are 
\eq{
\begin{array}{r c c c l}
z  &=&  l-s-t-u-v-w-x-y-10 &,&  \\
z' &=&  m-s'-t'-u'-v'-w'-x'-y' &,&  \\
z'\!' &=&  n-s'\!'-t'\!' -u'\!'-v'\!'-w'\!'-x'\!'-y'\!' &,& \\
\end{array}
}
and indices in coefficient $a_{r\!,r'\!,r'\!'}$ are
\begin{eqnarray}
r  &=&  \frac{1}{2}  (l-2s-2t-2u-10), \\
r' &=&  \frac{1}{2}  (m-2s'-2t'-2u'), \\
r'' &=&  \frac{1}{2}  (n-2s'\!'-2t'\!' -2u'\!').
\end{eqnarray}

We implemented the recursive equation numerically, and the resulting coefficients for several initial numbers of atoms are presented in Tab.~(\ref{tab:listIsomersRing}).
\begin{table}
    \centering
    \begin{tabular}{l | c  |l| c }
    $l,m,n$   & Benzene  & $ l,m,n $&  Naphthalene  \\\hline 
    $ 6,0,6 $ &  $  1 $     & $10,0,8    $ &  $ 1$ \\
    $ 6,1,6 $ &  $  1 $     & $10,1,8    $ &  $ 2$  \\
    $ 6,2,6 $ &  $  3 $     & $10,2,8    $ &  $ 10$  \\
    $ 6,3,6 $ &  $  3 $     & $10,3,8    $ &  $ 14$ \\
    $ 7,0,8 $ &  $  1 $     & $11,0,10   $ &  $ 2$ \\
    $ 7,1,6 $ &  $  1 $     & $11,1,8    $ &  $ 2$   \\
    $ 7,1,8 $ &  $  4 $     & $11,1,10   $ &  $ 16$   \\
    $ 7,2,6 $ &  $  4 $     & $11,2,8    $ &  $ 16$  \\
    $ 7,2,8 $ &  $  10$     & $11,2,10   $ &  $ 58$   \\
    $ 7,3,6 $ &  $  9 $     & $11,3,8    $ &  $ 56$  \\
    $ 7,3,8 $ &  $  16$     & $11,3,10   $ &  $ 128$   \\
    $ 8,0,6 $ &  $  1 $     & $12,3,10    $ &  $ 794 $  \\
    \vdots     & \vdots  &    \vdots     &  \vdots \\
   $ 10,0,8 $ &  $  7    $  & $ 14,2,10 $ &  $ 2674 $  \\
   $ 10,1,6 $ &  $  13   $  & $ 14,2,12  $ &  $ 7692 $  \\
   $ 10,1,8 $ &  $  82  $  & $ 14,2,14 $ &  $ 11184 $  \\
   $ 10,2,6 $ &  $  51   $  & $ 14,3,8  $ &  $ 1224 $  \\
   $ 10,2,8 $ &  $  389  $  & $ 14,3,10 $ &  $ 10288 $  \\
   $ 10,2,10 $ &  $  1117  $  & $ 14,3,12  $ &  $ 30432 $  \\
   $ 10,2,12 $ &  $  1520 $  & $ 14,3,14 $ &  $ 41456 $    \\
    \vdots     & \vdots  &     \vdots     &  \vdots \\
    \end{tabular}
    \caption{The number of isomers for organic molecules containing one ring (benzene) and two stuck rings (naphthalene) with $l,m,$ and $n$ carbon, oxygen, and hydrogen atoms.}
    \label{tab:listIsomersRing}
\end{table}
This procedure can be extended to a wide range of polycyclic structures, such as tricyclic, tetracyclic, and even more varieties of fused rings.


\section{Discussion} 
%
%
In this paper, we extended counting theory for chemical graphs to accommodate multiple bonds and account for different species of atoms, which necessitated the use of multivariate generating functions. We illustrated this approach to chemical counting with the example of two classes of ringlike molecules growing from benzene and naphthalene roots, respectively.  

The approach presented here could easily be extended to unrooted trees, structures with other roots, and even structures containing further isolated rings. A more interesting and perhaps more difficult challenge is to decide which structures to count. Clearly several structures allowed by the basic bonding rules of chemistry are extremely short-lived in solution and therefore unlikely to encountered in nature. Other cyclic structures may be allowed by bonding rules but geometrically impossible. Furthermore, compunds such as diamond and graphene are not only allowed by chemical rules and geometry but also highly stable. However, for typical environmental application where we seek to explore the potential molecular diversity in soils or seawater, we would not want to count such configurations.  

Main complications that are on the horizon in this area of are thus not purely mathematical. As we gain the mathematical ability to count more and more complex structures, decisions need to be made which classes of structures are sensible to count within the context of the application.

The need for modeling decision highlights a need to bring chemical counting from combinatorial graph theory into network science, where generating function calculations are routinely combined with modeling decisions and physics approximations. Using such approximation methods it would, for instance, be a promising next step to study the scaling behavior of molecular diversity. We hope that the advances presented here, and also our introduction to the topic will facilitate such studies in the future.


\bibliographystyle{plain}
\bibliography{REF}

\end{document}